\documentclass[12pt]{iopart}

\usepackage{graphicx} 
\begin{document}

\title[The Heat Distribution in a Logarithm Potential]{The Heat Distribution in a Logarithm Potential}

\author{Pedro V. Paraguassú}

\address{Departamento de F\'{i}sica, Pontif\'{i}cia Universidade Cat\'{o}lica\\ 22452-970, Rio de Janeiro, Brazil}

\author{Welles A.~M. Morgado}
\address{Departamento de F\'{i}sica, Pontif\'{i}cia Universidade Cat\'{o}lica\\ 22452-970, Rio de Janeiro, Brazil\\ and National  Institute of Science and Technology for Complex Systems}

\ead{paraguassu@aluno.puc-rio.br}
\vspace{10pt}
\begin{indented}
\item[]
\end{indented}

\begin{abstract}
All statistical information about the heat can be obtained with the probability distribution of the heat functional. This paper derives analytically the expression for the distribution of the heat, through path integral, for a diffusive system in a logarithm potential. We apply the found distribution to the first passage problem and find unexpected results for the reversibility of the distribution, giving a fluctuation theorem under specific conditions of the strength parameters.
\end{abstract}

%
%
%
%
%

\section{Introduction}
With the fast development of Stochastic Thermodynamics  in the past two decades, fluctuations in thermodynamic quantities, such as heat, work, and entropy, for diffusive systems, have been extensively investigated~\cite{sekimoto2010stochastic,seifert2012stochastic,ryabov2015stochastic}. These fluctuations can be studied through the functional distribution~\cite{taniguchi2007onsager,chatterjee2010exact}, or the characteristic function~\cite{Imparato2007,Imparato2008}. Exact and analytical results for heat distributions were first presented only through the generic use of path integrals, whereas  in~\cite{chatterjee2010exact} the distribution of heat, for a diffusive system under the action of a harmonic potential, is obtained for the first time. Since then, several results have been obtained for the distribution of heat in cases where the force is of the linear type \cite{chatterjee2010exact,chatterjee2011single,ghosal2016distribution,goswami2019heat}. As far as we know, analytical results for heat are only possible for free particles and linear forces. In the present work, we try to extend this rather short list by studying a diffusive system where the force is strongly non-linear \cite{bray2000random}, derived from a logarithmic potential.

Diffusive systems with logarithmic potential are found in the literature, and describe quite distinct phenomena phenomena such as 2D XY model with Kosterlitz-Thouless phase transition~\cite{bray2000random}, DNA bubble denaturation~\cite{fogedby2007dna}, sleep-walk dynamics~\cite{lo2002dynamics}, charged Brownian particle movement~\cite{liboff1966brownian}, and interaction models between poly-electrolyte polymers~\cite{manning1969limiting}. These two last examples have the same origin of the interaction, which is the Coulomb electrical potential of a charged line and serves to illustrate the present work. 

Recently, when including external protocols in the system, significant attention has been give to the work distributions, where analytical and numerical results were obtained~\cite{ryabov2013work,Holubec2015,Holubec2015a}, providing a broad characterization of the distribution under different protocols. Here, we try to investigate in another direction, studying the heat, which is an important thermodynamic quantity if we are interested in Brownian machines. Another recent study is the first passage problem with protocols in~\cite{ryabov2015brownian}. The first passage problem, consist of finding the distribution of the first time that the particles hit an absorbing boundary condition \cite{bray2000random}. Having a time probability distribution for this first time, we are able to find the average time behavior of functionals. We use this to see the averaged behavior of the heat through this process.

Having the distribution of a functional, one may wonder whether it obeys any fluctuation theorem~\cite{searles1999fluctuation,jarzynski1997nonequilibrium,crooks1999entropy,seifert2005entropy}. Fluctuation theorem is a link between the distribution and its reverse process. Today, we already know an unified fluctuation theorem for entropy \cite{seifert2012stochastic,chernyak2006path,sughiyama2011extended}. For heat, a theorem is only possible for specific conditions \cite{hatano2001steady,van2003extension,van2004extended,speck2005integral}. Inquisitively, we use the derived heat distribution to look for fluctuation theorems, finding unexpected results due to the non-trivial form of the distribution.

In the present work, we use path integral formalism~\cite{chaichian2018path,wio2013path,cugliandolo2017rules} to obtain the heat distribution for a logarithmic potential exactly, and analytically. We apply it to the first pass problem by taking a time average over the distribution, and we investigate the validity of fluctuation theorems for the distribution found.

In section II, we present the stochastic thermodynamics of the model, defining the heat functional with the Langevin equation. In section III, we solve for the heat distribution using path integrals. In section IV, we work out a simple application of the heat distribution in the first passage problem. In section V, we investigate the role of fluctuation theorem for the heat distribution. We finish in Section VI, with a discussion. In the  Appendix, we perform the calculation of the path integral.

\section{Stochastic Thermodynamics in Logarithm Potential}
We start with the overdamped Langevin equation, 
\begin{equation}
    \dot{x}(\tau)=-\frac{k}{x(\tau)}+\zeta(\tau)\label{langevin}
\end{equation}
This describes an overdamped Brownian particle suffering the action of an attractive  force
\[
f(x)=-\frac{k}{x}=-\partial_x U(x),
\] 
with $k>0$, which is derived from a logarithm potential $U(x)=k\ln{x}$, and a thermal force, which is a Gaussian white noise of zero average, obeying
\begin{equation}
    \langle \zeta(\tau)\zeta(\tau')\rangle=2D\,\delta(\tau-\tau'),
\end{equation} 
where $k$ and $D$ are the strength parameters of the problem representing, respectively, the force of the potential and the force of the reservoir. The dynamics of the particle occurs in a time interval $\tau\in[0,t]$, and is restrict for $x(\tau)>0$ with boundary conditions: $x(0)=x_0$ and $x(t)=x_t$. We assume a perfect absorbing boundary condition for the particle distribution in  $x(t^{\prime})=0$, it is $P(x=0,t^{\prime})=0$. The Langevin Eq.~\ref{langevin} is also known in the literature as the Bessel Process \cite{karlin2014first}.

The physical character of the force in  Eq.~\ref{langevin}, can be understood as the electrostatic force between an uniformly charged line  and an ion  \cite{zangwill2013modern}. Instead of a force of the type $\propto 1/r^2$, a charged line creates a Coulomb force  $\propto 1/r$. Here, the strength of the potential $k$ is associated with the charges of the particle and the line, and since the force is attractive, the charges must be  opposite. A realistic example, where thermal fluctuations take place, is the interaction of a polyelectrolyte and an ion, with opposite charges \cite{manning1969limiting}. While the ion is still diffusing, it has not yet been captured by the attractive force of the polymer. When the ion hits the polymer it gets stuck, hence the first passage occurs at that instant. This first passage time has a distribution \cite{bray2000random} that more fully characterizes the process, contrasting with the survival probability, which is the probability for the time that the particle still hasn't hit the polymer. In section V, we apply these concepts to the heat distribution.

According to Sekimoto \cite{sekimoto2010stochastic}, the total heat exchange between the particle and the reservoir in the interval $\tau\in[0,t]$ is defined as
\begin{equation}
    Q[x]=-\int_0^t \left(\dot{x}(\tau)-\zeta(\tau)\right)\frac{dx}{d\tau}d\tau=\int_0^t\frac{k}{x}\frac{dx}{d\tau}d\tau=k\ln{\left(\frac{x_t}{x_0}\right)}\label{heat}
\end{equation}
where we use the Langevin Eq.~\ref{langevin}. The above equation expresses the balance of energy between the particle and the reservoir, and can be seen as the first law of stochastic thermodynamics, since $\Delta U(x)=k\ln{\left(x_t/x_0\right)}$, and then the first law reads \[Q[x]=\Delta U.\] Notice that no work is done in the system since we have not assigned any work protocol. Nevertheless, we shall see in the next section that this case still gives a non trivial result for this functional. 

\section{Distribution for the Heat Functional}
In Stochastic Thermodynamics \cite{seifert2012stochastic,sekimoto2010stochastic}, heat is a functional of the particle's trajectory $x(t)$~\cite{Crooks2000}. Due to the dependence on the stochastic trajectory, heat is a stochastic variable, having an associated distribution $P(Q,t)$ that contains all information about the statistical properties of the heat. To find an exact expression for $P(Q,t)$ we will use path integral techniques \cite{wio2013path,taniguchi2007onsager}, where the distribution is given by the average 
\begin{equation}
   \fl P(Q,t)=\langle \delta(Q-Q[x]) \rangle =\int dx_t \int dx_0 P_0(x_0) \int_{x_t,x_0} Dx\; e^{\mathcal{A}[x(t)]} \delta(Q-Q[x]).\label{heatdist}
\end{equation}
In principle, we have to solve the path integral, defined in the above equation, by taking into account the constraint imposed by the Dirac delta. For our system, the heat $Q[x]$ given by Eq.~\ref{heat} only depends on the boundary conditions $x_0\rightarrow x_t$, then we just take the Dirac delta out of the path integral and use it to solve one of the remaining integrals. Nevertheless, we still need to solve the path integral
\begin{equation}
    \int_{x_t,x_0} Dx e^{\mathcal{A}[x(t)]} = P[x_t,t|x_0,0], \label{pathintegral}
\end{equation}
which is the conditional probability for the Langevin equation given in Eq.~\ref{langevin}, and $\mathcal{A}[x]$ is the stochastic action \cite{wio2013path} defined in Eq.~\ref{action2}. The solution of the path integral in Eq.~\ref{pathintegral} is already known for a more general case in \cite{giampaoli1999exact} and we give a complete derivation in appendix A.

\begin{figure}
    \centering
    \includegraphics[width=15cm]{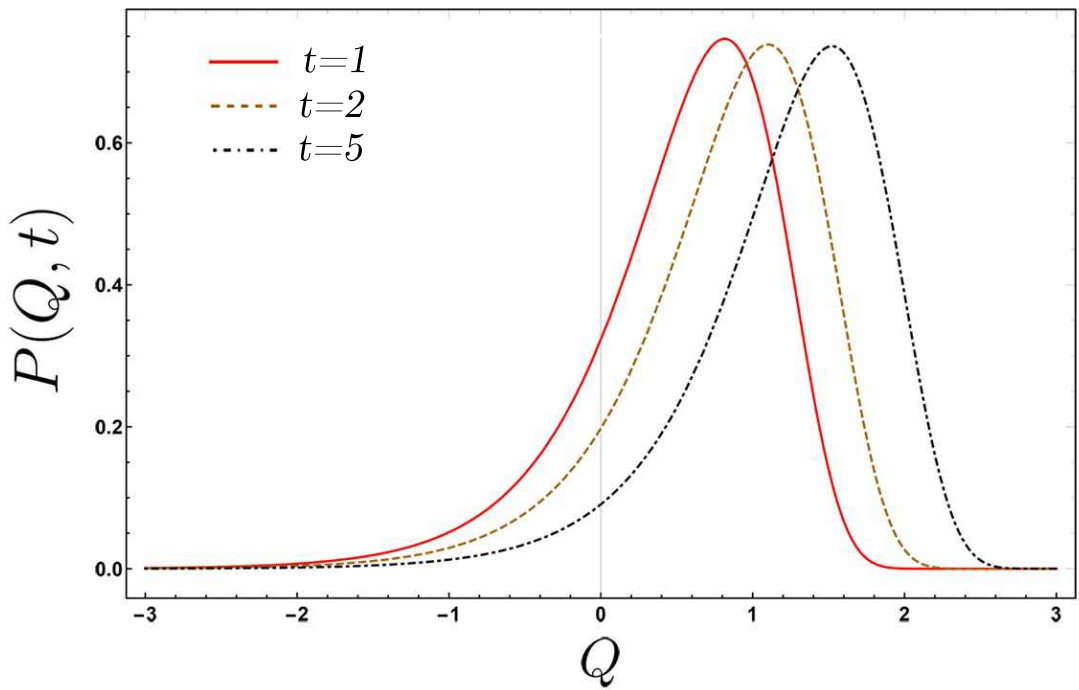}
    \caption{The Heat distribution with initial delta probability for $k=1,D=1,x_i=1$ with three different times $t=1,t=2,t=5$.}
    \label{distheatplot}
\end{figure}

With the simplification given by the model, we have a more simple formula for the distribution $P(Q,t)$, that can be further simplified to 
\begin{equation}
    P(Q,t)=\int dx_t \; \delta\left(Q-k\ln{\frac{x_t}{x_0}}\right) \int dx_0 P_0(x_0) P[x_t,t|x_0,0],
\end{equation}
where the Dirac delta can be rewritten in a more convenient way by
\begin{equation}
    \delta\left(Q-k\ln{\frac{x_t}{x_0}}\right)=\frac{x_0e^{Q/k}}{k} \delta\left(x_0e^{Q/k}-x_t\right).
\end{equation}
Integrating  over $x_t$, we find 
\begin{equation}
    P(Q,t)=\frac{e^{Q/k}}{k}\int dx_0 x_0P_0(x_0) P[x_0e^{Q/k},t|x_0,0] .
\end{equation} 
We reduce the formula to just one integral over $x_0$ with the initial distribution $P_0(x_0)$. 
For the delta distribution as the initial distribution $P_0(x_0)=\delta(x_0-x_i)$, we can find an exact and analytical result for the heat distribution. Using the Dirac delta property, we solve the integral and find
\begin{equation}
    P(Q,t)=\frac{x_ie^{Q/k}}{k}P\left(x_ie^{Q/k},t|x_i,0\right).
\end{equation}
Using the conditional probability given in Eq.~\ref{transitionalprob}, we have the explicit formula
\begin{equation}
   \fl P(Q,t)=\frac{1}{4Dkt}I_\kappa\left(\frac{e^{Q/k}x_i^2}{2Dt}\right)\left(e^{Q/k}x_i\right)^{\kappa+2}x_i^\kappa\exp\left(-\frac{1}{4Dt}\left(e^{2Q/k}+1\right)x_i^2\right), \label{Heatdistdelta}
\end{equation}
which is valid to all times, and $x_i>0$. $I_\kappa$ is the Modified Bessel function of the first kind, and $\kappa=\frac{1}{2}(k/D-1)$ is a parameter defined in the appendix A. The exact expression for the heat distribution in Eq.~\ref{Heatdistdelta} is the main result of this work. The plot of Eq.~\ref{Heatdistdelta} is given in figure~\ref{distheatplot}. We can see that the most probable values are in the positive $Q>0$ region. As time passes, only positive values for heat are allowed. The particle is predominantly gaining energy from the reservoir.

For the asymptotic behavior, $t\rightarrow \infty$, the argument of the Bessel function in Eq.~\ref{Heatdistdelta} is small, and then the expansion for small values of the Bessel function yields  
\[
I_\kappa(z)\propto\frac{z^\kappa}{2^\kappa\Gamma(\kappa+1)},
\] 
showing a crescent exponential tail for $Q$. The normalization for this distribution can be checked numerically, and is given by $\int_\infty^\infty dQP(Q)=1$ as expected.

\section{Application to the first passage problem}

\begin{figure}
    \centering
    \includegraphics[width=17cm]{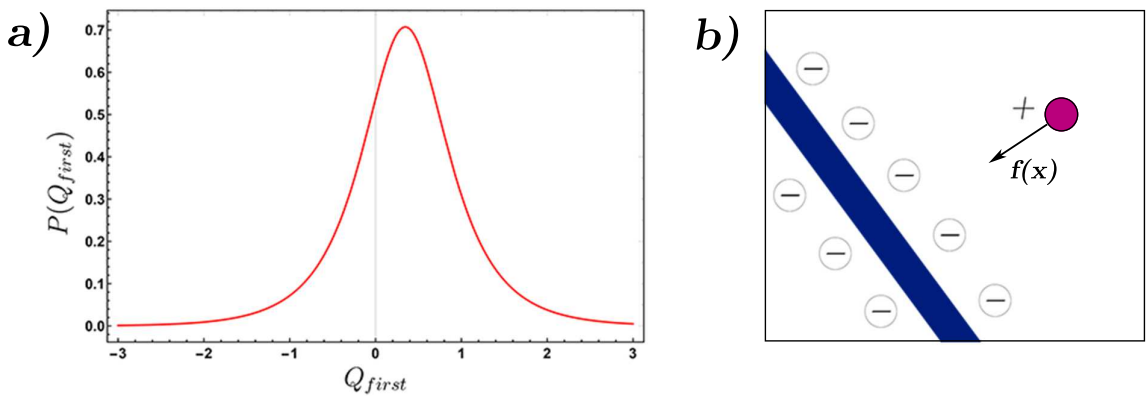}
    \caption{(a) Distribution for the first time heat $Q_{first}$ for $k=1,D=1,x_i=1$. (b) Pictorial representation of a ion with positive charge (purple circle) and a polyelectrolyte with negative charge (blue line).}
    \label{heatfpp}
\end{figure}

To illustrate the phenomena, we consider the case of an interaction of a polyelectrolyte, which is a charged polymer, and an ion \cite{manning1969limiting}. In a system with a polyelectrolyte and an ion with opposite charge, see figure~\ref{heatfpp} (b), the Coulomb interaction between them is attractive through a logarithm potential \cite{ryabov2015brownian}. Therefore the Langevin Eq.~\ref{langevin} of our model, can be used to describe this interaction.

This kind of system exhibits the first passage problem, which corresponds to the first time that the ion hits the polymer. The probability for the first time $P_1(t)$ is given by \cite{bray2000random},
\begin{equation}
    P_1(t)=\frac{1}{\Gamma(\kappa+1)}\frac{4D}{x_i^2}\left(\frac{x_i^2}{4Dt}\right)^{\kappa+2}\exp\left(\frac{-x_i^2}{4Dt}\right).\label{firsttime}
\end{equation}
The heat distribution found in the previous section is valid to all times $t$. With $P(Q,t)$ we can ask: What is the statistical behavior of the heat like in a first passage event? To answer, we can average the heat distribution in Eq.~\ref{Heatdistdelta} in time with the first time probability in Eq.~\ref{firsttime}, that is
\begin{equation}
    P(Q_{first})=\int_0^\infty P_1(t) P(Q,t) dt.
\end{equation}
However, we can't solve it analytically and we need to solve numerically. Using Wolfram Mathematica \cite{Mathematica} we can plot the distribution for the first passage heat, which is shown in figure~\ref{heatfpp} (a).

When the ion moves towards the polyelectrolyte, heat is exchanged with the reservoir.
The distribution $P(Q_{first})$ shown in figure~\ref{heatfpp}(a) gives the typical heat for the first passage problem.

\section{Fluctuation Theorem}

To study the path reversibility hidden in a probability distribution, we can calculate the ratio between the distribution and the inverse distribution. In some cases this ratio gives a fluctuation theorem \cite{seifert2012stochastic,seifert2005entropy,searles1999fluctuation,van2003stationary}. The heat is connected to the entropy production through $Q[x]/T=S[x]$. For steady states, it is well know that the entropy production obeys a fluctuation theorem \cite{seifert2005entropy}. It is then desired to see if the distribution found in the previous section satisfies some kind of fluctuation theorem. 

The distribution of $P(Q)$ is not trivial, in addition to exponentials, we also have a Bessel function $I_\kappa$ that depends on the exponential of $Q$, this lets us think that, in principle, there is no standard fluctuation theorem, as is shown below
\begin{equation}
    \frac{P(Q)}{P(-Q)}=\frac{I_\kappa\left(\frac{e^{Q/k} x_i^2}{2 D t}\right) }{I_\kappa\left(\frac{e^{-\frac{Q}{k}} x_i^2}{2 D t}\right)}\exp \left(-\frac{x_i^2 \sinh \left(\frac{2 Q}{k}\right)}{2 D t}+\frac{Q (D+k)}{D k}+\frac{2 Q}{k}\right). \label{fluct1}
\end{equation}
It is straightforward to see that the distribution for the heat does not exhibit a standard  fluctuation theorem form. However, in the asymptotic limit, when $t\rightarrow\infty$, the argument in the Bessel function becomes small, as noted in the previous section, so we can derive a power law from the Bessel function $I_\kappa$. Therefore, we find the measure of the reversibility of $Q$
\begin{equation}
    \ln\left(\frac{P(Q,t)}{P(-Q,t)}\right)\asymp \frac{4}{k}(1+\kappa)Q\;, \label{fluct2}
\end{equation}
where $\asymp$ means that the equality holds only for $t\rightarrow\infty$. In Eq.\ref{fluct2} we don't have a fluctuation theorem unless 
\begin{equation}
    k=\frac{D}{-1+D/2}.\label{condi}
\end{equation}
If the above condition between the strength of the potential and the strength of the bath is satisfied, a Stationary State Fluctuation Theorem (SSFT) \cite{van2003stationary} is satisfied for the Heat.

\begin{figure}
    \centering
    \includegraphics{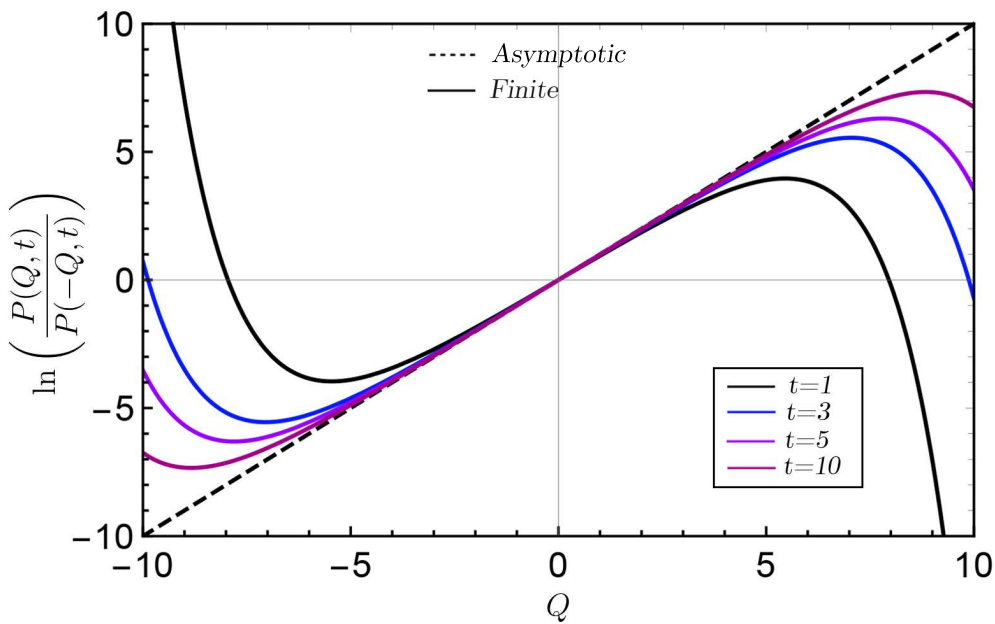}
    \caption{Log ratio probability in the asymptotic limit (dashed line) and finite values of time $t=1,3,5,10$ (solid lines) for $k=3,D=6$ satisfying $k=\frac{D}{-1+D/2}$. 
    Observe that, as time increases, the log-ratio shall favor less and less the high negative values over their opposite positive ones. Actually, the picture shows only their probability ratio (which are, in fact, both small).}
    \label{fluctfig}
\end{figure}

The plot of the two versions of the ratio, Eq.\ref{fluct1} and \ref{fluct2}, satisfying the condition in Eq.\ref{condi} are in Figure \ref{fluctfig}, which shows the agreement between the two equations in a finite region. 

\section{Discussion and Conclusion}
In the present work, we studied the distribution of heat for a diffusive system in a logarithmic potential. We find analytical, and exact, results for the heat distribution through the use of path integrals enlarging the list of few exact results in this field. Here, we limit ourselves to the case where the interaction is of the attractive type, where the first pass problem occurs. We then use the distribution found to obtain the average heat behavior during this process. To illustrate, we use the example of an ion and a charged polymer, which is carried out experimentally. Finally, we investigate the validity of the fluctuation theorem for heat through the log-ratio probability where we find a fluctuation theorem for heat only for long times and specific values of the strength parameters. 

The distribution of the heat in Eq.~\ref{Heatdistdelta}, obtained analytically through path integral technique, is the main result of this work. The expression obtained is a non-trivial distribution, with an exponential dependence in the argument of a modified Bessel function of the first kind, and an exponential in the argument of an exponential. These non-trivial dependencies come from the conditional probability of the Bessel process and the logarithm dependence in the boundary conditions of the heat functional in Eq.~\ref{heat}. The plot of the distribution gives us more physical insight. For fixed values of the strength parameters we can see the time evolution of the distribution; as time passes, negative values of the heat become less probable while positive values become most probable. Then while the particle survives, the gaining of energy from the reservoir is predominant. This can be understood noting that the thermal force has to overcome the attractive force, making it more probable to the survival particle to receive energy from the reservoir.

One of the applications of the work is in the first passage problem of a charged polymer and an ion with opposite charges. Using the probability distribution for the first time that the ion hits the polymer we average the heat distribution, finding the average heat for this process, shown in figure \ref{heatfpp}. We called $P(Q_{first})$ the averaged distribution, which gives the average statistical behavior of the heat for the process. $P(Q_{first})$ has similar statistical properties of $P(Q,t)$, where the most probable values are for positive heat. However, the shape of the distribution becomes more smooth, having a broader area.

Beyond the exact expression of the heat distribution and the application in the first passage problem, we investigate the role of fluctuation theorems for the heat distribution. Having a limited domain for the position of the particle $x(t)>0$, and the delta distribution as the initial distribution, a fluctuation theorem is not expected. However, the log ratio of the heat distribution, Eq.\ref{fluct2}, gives a fluctuation theorem for long times and a specific constraint, Eq.~\ref{condi}, between the strength parameters. As far we know, there is no physical consequence if the strength constraint is satisfied, we interpret this result as a mathematical aspect of the distribution. The fluctuation theorem found is an SSFT \cite{van2003stationary} for the heat, there is a connection between this theorem and fluctuation theorem for the entropy \cite{seifert2005entropy} since the entropy production in the reservoir has a direct relation with the heat functional. The log-ratio plotted in figure \ref{fluctfig}, shows the evolution towards the fluctuation theorem in the asymptotic limit $t\rightarrow\infty$. We can justify (that the fluctuation theorem is only valid asymptotically) in the use of a non-thermalized initial distribution. To be sure of this justification, an investigation using the initial thermalized condition, i.e., the Boltzmann distribution, is necessary and may be the subject of future work. Although the fluctuation theorem was not the initial motivation for the present work, the investigation in this direction found unexpected results.

Some questions are left behind and can serve as an extension of the present work; what is the effect of having an initial thermal condition? and how does heat behave when the interaction is strongly repulsive? where the first pass problem does not occur.

\ack

This work is supported by the Brazilian agencies CAPES and CNPq. PVP would like to thank CNPq for his current fellowship.   This study was financed in part by Coordena\c c\~ ao de Aperfei\c coamento de Pessoal de N\' ivel Superior - Brasil (CAPES) - Finance Code 001.

\appendix

\section{Path Integral for the Conditional probability}
The stochastic action $\mathcal{A}[x]$ that appears in Eq.~\ref{pathintegral} is defined by \cite{chaichian2018path,wio2013path,onsager1953fluctuations}
\begin{equation}
   \fl \mathcal{A}[x]\rightarrow\mathcal{A}[\dot{x},x,t]=-\frac{1}{4D}\int_0^t\left[\dot{x}^2+\left(k^2-2Dk\right)\frac{1}{x(s)^2}\right]ds +\frac{k}{2D}\ln\left(\frac{x_t}{x_0}\right) \label{action2}
\end{equation}
where we use the Stratonovich convention and the last term can be take off of the path integral, so we define the propagator $K[x_t,t|x_0,0]$ as
\begin{equation}
    P[x_t,t|x_0,0]=e^{\frac{k}{2D}\ln\left(\frac{x_t}{x_0}\right)}K[x_t,t|x_0,0]
\end{equation}
where $K(x_t,t|x_0,0)$ is now just the path integral
\begin{equation}
   \fl K[x_t,t|x_0,0]=\int_{x(0)=x_0}^{x(t)=x} \mathcal{D}x\;\exp\left[-\frac{1}{4D}\int_0^t\left(\dot{x}^2+\left(k^2-2Dk\right)\frac{1}{x(s)^2}\right)ds\right]. \label{propagatorK}
\end{equation}{}

In order to solve the path integral in Eq. \ref{propagatorK}, we use the approximation by piecewise linear functions described in \cite{chaichian2018path}, this is essentially the same approach done in \cite{wio2013path,giampaoli1999exact}. We discretize the path $x(s)$ in piecewise linear functions $x_i=x(s_i)$ and the time $s_i=i\epsilon$ in $N$ pieces $i=1,...,N$, with $N\epsilon=t$ where $s_N=t$ is the final time. At the end of the calculation, we take the limit $N\rightarrow \infty$. The action $\mathcal{A}[x]$ in Eq. \ref{pathintegral} in the discretized form is
\begin{equation}
    \mathcal{A}_N=-\frac{1}{4D}\sum_{i=0}^{N-1} \frac{(x_{i+1}^2+x_i^2)}{\epsilon}+\sum_{i=0}^{N-1}\left[ \frac{x_ix_{i+1}}{2D\epsilon} -  \frac{1}{2}\frac{(\kappa^2-\frac{1}{4})2D\epsilon}{x_ix_{i+1}}    \right]\label{actiondiscre}
\end{equation}
where we define the constant ${\left( {\kappa }^{2 }-\frac{1 }{4 }\right) }=\frac{{k }^{2 }-2 D k }{4 {D }^{2 }}\rightarrow \kappa=\frac{1}{2}\left(\frac{k}{D}-1\right)$ for later convenience because defining $\kappa$ allow us to use the following expansion of the exponential of the last term in Eq. \ref{actiondiscre} in terms of modified Bessel functions $I_\kappa(x)$\cite{watson1962treatise}
\begin{equation}
   \fl \exp\left( \frac{x_ix_{i+1}}{\epsilon} -  \frac{1}{2}\left(\kappa^2-\frac{1}{4}\right)\frac{\epsilon}{x_ix_{i+1}}  +\mathcal{O}(\epsilon^2)  \right)\sqrt{\frac{2\pi x_ix_{i+1}}{\epsilon}}I_\kappa\left(\frac{x_ix_{i+1}}{\epsilon}\right),\label{besseltransform}
\end{equation}
we can simplify our Path Integral. This approximation works, because we want to take the limit $N\rightarrow\infty$, or equivalent, $\epsilon\rightarrow0$, and then, terms of order $\mathcal{O}(\epsilon^2)$ can be ignored. This is the crucial step to evaluate the path integral in Eq. \ref{pathintegral}. Since the beginning, we don't have a Gaussian integral, an exact solution is not available unless we transform Eq. \ref{besseltransform}.

The next step is to consider the Path Integral as multiple integrals in the $x$-coordinate. Since we are dealing only with the cases where $x>0$, the integrals over $x$'s are in the range $0\rightarrow\infty$, different from usual. Defining the $N$-integral $K_N$ as 
\begin{equation}
    K_N= \left(\frac{1}{4\pi D \epsilon}\right)^{N/2}\int_0^\infty dx_0 \;... \int_0^\infty dx_N \exp{\mathcal{A_N}}
\end{equation}{}
where the first term comes from the integration measure $\mathcal{D}x$, we can use the definition in Eq. \ref{actiondiscre} and the identity in Eq. \ref{besseltransform} 
\begin{equation}
  \fl  \eqalign{ K_N = \left(\frac{1}{4\pi D \epsilon}\right)^{N/2}\int_0^\infty dx_0 \;... \int_0^\infty dx_N \exp\left(-\frac{1}{4D}\sum_{i=0}^{N-1} \frac{(x_{i+1}^2+x_i^2)}{\epsilon}\right)\\\times\prod_{i=0}^{N-1}\sqrt{\frac{2\pi x_ix_{i+1}}{2D\epsilon}}I_\kappa\left(\frac{x_ix_{i+1}}{2D\epsilon}\right).}\label{pathint3}
\end{equation}
The first term in the integral of Eq. \ref{pathint3} can be decomposed in
\begin{equation}
    -\frac{1}{4D}\sum_{i=0}^{N-1} \frac{(x_{i+1}^2+x_i^2)}{\epsilon}=-\frac{x_0^2+x_N^2}{4D\epsilon}-\sum_{i=1}^{N-1}\frac{x_i^2}{2D\epsilon},
\end{equation}{}
and the term in the square root
\begin{equation}
    \prod_{i=0}^{N-1}\sqrt{\frac{2\pi x_ix_{i+1}}{2D\epsilon}}\rightarrow\sqrt{x_0x_N}\prod_{i=1}^{N-1}\sqrt{\frac{2\pi }{2D\epsilon}}x_i=\left(\frac{1}{2D\epsilon}\right)^{N/2}\sqrt{2\pi}^N\prod_{i=1}^{N-1}x_i
\end{equation}{}

Then, the integral in Eq. \ref{pathint3} can be rewritten as
\begin{equation}
   \fl  K_N=\mathcal{N}\sqrt{x_0x_N}\int_0^\infty  \;... \int_0^\infty \prod_{i=1}^{N-1} dx_i\; e^{-\frac{1}{2D\epsilon}x_i^2}I_\kappa\left(\frac{x_0x_{1}}{2D\epsilon}\right)I_\kappa\left(\frac{x_ix_{i+1}}{2D\epsilon}\right)x_i,
\end{equation}
where, $\mathcal{N}$ is a normalization factor defined by $\mathcal{N}={{\left( \frac{1 }{2 D \epsilon }\right) }}^{N }e^{-\frac{x_0^2+x_N^2}{4D\epsilon}} $. Note that the multiplicand now runs from $i=1$ to $i=N-1$, instead of starting in $i=0$. We still need to integrate over all $x$'s, and to do this we use the property of Bessel Functions \cite{watson1962treatise}
\begin{equation}
    \int_0^\infty e^{-\alpha y^2}I_\kappa(a_1y)I_\kappa(a_2y)ydy=\frac{1}{2\alpha}\exp\left(\frac{a_1^2+a_2^2}{4\alpha}\right)I_\kappa\left(\frac{a_1a_2}{2\alpha}\right)
\end{equation}{}
This integral is valid for $\alpha>0$ and $\kappa>-1$, which is our case. 

The first integral in $x_1$ can be computed as
\begin{equation}
   \fl \int_0^\infty e^{-\frac{1}{2D\epsilon}x_1^2}I_\kappa\left(\frac{x_0x_{1}}{2D\epsilon}\right)I_\kappa\left(\frac{x_1x_{2}}{2D\epsilon}\right)x_1\;dx_1 = \frac{1 }{2 c }{e }^{{\left( {x _{0 }}^{2 }+{x _{2 }}^{2 }\right) }\frac{c }{4 }}I {\left( x _{0 }x _{2 }\frac{c }{2 }\right) },
\end{equation}
where we define $c\equiv \frac{1}{2D\epsilon}$ for later convenience. The next integral in $x_2$ will give, 
\begin{equation}
    \eqalign{ &\frac{1 }{2 c }{e }^{{x _{0 }}^{2 }\frac{c }{4 }}\int _{0 }^{\infty }{e }^{-c {\left( 1 -\frac{1 }{4 }\right) }{x _{2 }}^{2 }}I {\left( x _{0 }x _{2 }\frac{c }{2 }\right) }I {\left( x _{3 }x _{2 }c \right) }x _{2 }d x _{2 }\; =\\&\frac{1 }{2 c }\cdot \frac{1 }{2 c {\left( 1 -\frac{1 }{4 }\right) }}{e }^{\frac{{x _{0 }}^{2 }}{4 }c {\left( 1 +\frac{1 }{4 {\left( 1 -\frac{1 }{4 }\right) }}\right) }}I {\left( x _{0 }x _{3 }\frac{c }{2 }\frac{1 }{2 {\left( 1 -\frac{1 }{4 }\right) }}\right) }  }
\end{equation}{}
and then, after $N-1$ integrations, we arrive at
\begin{equation}
    \frac{1}{2}\prod_{j=1}^{N-1}\frac{1}{2c\gamma_j}I_\kappa\left(x_0x_N\frac{c}{2}\prod_{j=1}^{N-1}\frac{1}{2\gamma_j}\right)\exp\left(P_Nx_0^2+x_N^2\frac{c}{4\gamma_{N-1}}\right)
\end{equation}{}
where \[\gamma_0=1, \;\;\;\;\; \gamma_j=\left(1-\frac{1}{4\gamma_{j-1}}\right),\;\;\;\;
P _{N }=c\sum _{j =1 }^{N -1 }\frac{1 }{{4 }^{j }\gamma _{j }}{{\left( \prod _{k =1 }^{j -1 }\frac{1 }{\gamma _{k }}\; \right) }}^{2 }\; 
\]

Putting all pieces together, we have for $K_N$ (including the normalization constant)
\begin{equation}
    \eqalign{K_N=\sqrt{x_0x_N}\;\frac{c}{2}\prod_{j=1}^{N-1}\frac{1}{2\gamma_j}I_\kappa\left(x_0x_N\frac{c}{2}\prod_{j=1}^{N-1}\frac{1}{2\gamma_j}\right)\\\times\exp\left(x_0^2\left(P_N-\frac{c}{2}\right)+x_N^2\left(\frac{c}{4\gamma_{N-1}}-\frac{c}{2}\right)\right).}\label{Kn}
\end{equation}{}
To evaluate the path probability we need to take the limit $N\rightarrow\infty$. The first multiplicand that appears in the right side of Eq.\ref{Kn} and in the argument of the Bessel $I_\kappa$ can be calculate it as
\begin{equation}
    \frac{1}{2}\prod_{j=1}^{N-1}\frac{1}{2\gamma_j}=\frac{1}{2^N}\prod_{j=1}^{N-1}\frac{1}{\gamma_j}=\frac{1}{N} \;\;\;\Longrightarrow\;\;\frac{c}{N}=\frac{1}{2D t},\label{an}
\end{equation}
 where this equation gives $\prod_{j=1}^{N-1}\frac{1}{\gamma_j}=2^N/N$, and verification of this equation is just made by induction.
 
 The term $P_N$, which comes multiplied with $x_0$, needs a more careful approach, because has some divergences. To deal with it, we define $\gamma_j=\frac{Q_{j+1}}{2Q_j}$ such that
 \begin{equation}
     \gamma_j=1-\frac{1}{4\gamma_j}\;\;\rightarrow\;\;Q_{j+1}-2Q_{j}+Q_{j-1}=0\rightarrow \ddot{Q}=0,
 \end{equation}
where the last equation is the continuum limit of the discrete equation. The solution is just $Q(s)=s$ for $Q(0)=0$ as the initial condition. Backing to $P_N$, we can rewrite the term which comes multiplied with $x_0$
\begin{equation}
    P _{N }-\frac{1}{2}=-\frac{c }{2 }+ \frac{c }{2 }\prod _{k =1 }^{j -1 }\frac{1 }{2\gamma_{j }}\prod _{k =1 }^{j }\frac{1 }{2 \gamma _{j }}\; \;  
    =-\frac{c }{2 }+\frac{c }{2 }\sum _{j =1 }^{N -1 }\frac{{Q _{1 }}^{2 }}{Q _{j +1 }Q _{j }}.\;  \label{pn1}
\end{equation}
The last term in the continuum limit ($N\rightarrow\infty)$ is
\begin{equation}
        \frac{c }{2 }\sum _{j =1 }^{N -1 }\frac{{Q _{1 }}^{2 }}{Q _{j +1 }Q _{j }}\;  \rightarrow \lim_{\epsilon\rightarrow 0} \frac{1}{4D}\frac{Q _{1 }}{{\epsilon }^{2 }}\int _{\epsilon }^{t }\frac{1 }{Q {{\left( t ' \right) }}^{2 }}d t' \;\label{pn2} 
\end{equation}
We can rewrite the term outside the integral as
$\frac{Q_1-Q_0}{\epsilon}=\dot{Q}(0)=1,
$
because $Q_0=0$ is our initial condition. With Eq. \ref{pn2}, the continuum limit of Eq. \ref{pn1} is
\begin{eqnarray}
      \eqalign{\lim_{\epsilon\rightarrow\infty}P_N-\frac{c}{2}&= -\frac{1 }{4 D }{\left( \frac{1 }{\epsilon }-\int _{\epsilon }^{t }\frac{1 }{Q {{\left( t ' \right) }}^{2 }}d t' \; \right) } \\&=
       -\frac{1 }{4 D }{\left( \frac{1 }{\epsilon }-\int _{\epsilon }^{t }\frac{1 }{t'^{2 }}d t'\right)}
       \\&=-\frac{1}{4D t}.}\label{pn}
\end{eqnarray}{}

For the term multiplied with $x_N$, $\frac{c}{4\gamma_{N-1}}-\frac{c}{2}$, we just note that
$\frac{1}{4\gamma_{N-1}}=\frac{1}{2}\frac{Q_{N-1}}{Q_N},$ then the continuum limit is
\begin{equation}
   \fl \lim_{\epsilon\rightarrow0}\left(\frac{c}{4\gamma_{N-1}}-\frac{c}{4}\right)=\lim_{\epsilon\rightarrow0}\;-\frac{1 }{4 D Q _{N }}\frac{{\left( Q _{N +1 }-Q _{N }\right) }}{\epsilon }=-\frac{1}{2D}\frac{\dot{Q}(t)}{Q(t)}=-\frac{1}{4Dt}.\label{qn}
\end{equation}{}

Finally, using Eq. \ref{an}, \ref{pn} and \ref{qn} the propagator is
\begin{equation}
    K[x,t|x_0,0]=\lim_{N\rightarrow\infty}K_N=\sqrt{x_0x_t}\frac{1}{2D t}I_\kappa\left(\frac{x_0x_t}{2Dt}\right)e^{-\frac{1}{4D t}(x_t^2+x_0^2)},
\end{equation}
and then the conditional probability Eq.\ref{pathintegral} is
\begin{equation}
    P[x_t,t|x_0,0]=\frac{x_t^{\kappa+1}}{x_0^\kappa}\frac{1}{2D t}I_\kappa\left(\frac{x_0x_t}{2Dt}\right)e^{-\frac{1}{4D t}(x_t^2+x_0^2)}.\label{transitionalprob}
\end{equation}

\end{document}